\documentclass[12pt]{article}
\usepackage{epsfig}
\usepackage{graphicx}
\usepackage{rotating}
\usepackage{latexsym}
\usepackage{amsmath}
\usepackage{amssymb}
\usepackage{relsize}
\usepackage{geometry}
\usepackage{color}
\usepackage{bm}
\usepackage{slashed}

\def\beq{\begin{equation}}
\def\eeq{\end{equation}}
\def\beqn{\begin{eqnarray}}
\def\eeqn{\end{eqnarray}}
\def\beqn{\begin{eqnarray}}
\def\eeqn{\end{eqnarray}}

\def\ba{\beq\new\begin{array}{c}}
\def\ea{\end{array}\eeq}

\newcommand{\nonen}{${\mathcal N}=1$}
\newcommand{\ntwo}{${\mathcal N}=2$}

\newcommand{\ntwon}{${\mathcal N}=2$}
\newcommand{\ntwot}{${\mathcal N}= \left(2,2\right) $ }
\newcommand{\ntwoo}{${\mathcal N}= \left(0,2\right) $ }
\newcommand{\ntwoon}{${\mathcal N}= \left(0,2\right)$}

\newcommand{\ca}{{\mathcal A}}

\newcommand{\pt}{\partial}

\newcommand{\p}{\partial}
\newcommand{\wt}{\widetilde}
\newcommand{\ov}{\overline}
\newcommand{\mc}[1]{\mathcal{#1}}
\newcommand{\md}{\mathcal{D}}
\newcommand{\ml}{\mathcal{L}}
\newcommand{\mw}{\mathcal{W}}
\newcommand{\ma}{\mathcal{A}}

\newcommand{\lgr}{\left\lgroup}
\newcommand{\rgr}{\right\rgroup}

\newcommand{\LN}{\Lambda_\text{SU($N$)}}
\newcommand{\sunu}{{\rm SU($N$) $\times$ U(1) }}

\def\cfl {$\text{SU($N$)}_{\rm C+F}$ }

\newcommand{\AU}{\mc{A}^{\rm U(1)}}
\newcommand{\AN}{\mc{A}^\text{SU($N$)}}
\newcommand{\aU}{a^{\rm U(1)}}
\newcommand{\aN}{a^\text{SU($N$)}}

\newcommand{\baN}{\ov{a}{}^\text{SU($N$)}}

\newcommand{\nbar}{\ov{n}}
\newcommand{\nnbar}{n\ov{n}}
\newcommand{\muU}{\mu_\text{U}}

\newcommand{\cpn}{CP$^{N-1}$\,}

\newcommand{\qt}{\wt{q}}
\newcommand{\bq}{\ov{q}}
\newcommand{\bqt}{\overline{\widetilde{q}}}

\newcommand{\Tr}{\text{Tr}}
\newcommand{\Ts}{\text{Ts}}
\newcommand{\dm}{\hat{{\scriptstyle \Delta} m}}
\newcommand{\dmdag}{\hat{{\scriptstyle \Delta} m}{}^\dag}
\newcommand{\mhat}{\widehat{m}}
\newcommand{\deltam}{{\scriptstyle \Delta} m}
\newcommand{\nvac}{\vec{n}{}_\text{vac}}

\newcommand{\ie}{{\it i.e.}~}

\newcommand{\ansatz}{{\it ansatz} }

\begin{document}

\begin{titlepage}

\begin{flushright}
FTPI-MINN-13/29, UMN-TH-3301/13\\
\end{flushright}

\vspace{1.6cm}
\begin{center}
{\Large \bf  Twisted--Mass Potential \\
	on the Non--Abelian String World Sheet \\[2.4mm]
             Induced by Bulk Masses}
\end{center}

\vspace{2mm}

\begin{center}

 {\large
 \bf   Pavel A.~Bolokhov$^{\,a}$,  Mikhail Shifman$^{\,b}$ and \bf Alexei Yung$^{\,\,b,c}$}
\end {center}

\begin{center}

$^a${\it Theoretical Physics Department, St.Petersburg State University, Ulyanovskaya~1, 
	 Peterhof, St.Petersburg, 198504, Russia}\\
$^b${\it  William I. Fine Theoretical Physics Institute,
University of Minnesota,
Minneapolis, MN 55455, USA}\\
$^c${\it Petersburg Nuclear Physics Institute, Gatchina, St. Petersburg
188300, Russia
}
\end{center}

\vspace{1cm}
\begin{center}
{\large\bf Abstract}
\end{center}

	We derive  the twisted--mass potential in $ {\cal N}=(2,2) $ \cpn theory 
	on the world sheet of the non--Abelian string from the bulk \ntwo\, theory with massive (s)quarks
	by determining the profile functions of the adjoint fields.
	Although this potential was indirectly found some time ago, this is the first  
	direct derivation from the bulk.

	As an application of the adjoint field profiles, we compute and confirm the $ |\, \mu \sigma \,| $ potential 
	(where $\sigma$ is a scalar field in the gauge supermultiplet),
	which arises in the effective two--dimensional theory on the string 
	due to the supersymmetry breaking bulk mass term $ \mu\, \ca^2 $ for the adjoint matter.

\vspace{2cm}

\end{titlepage}

\section{Introduction}
\setcounter{equation}{0}

	Phenomena on the non-Abelian flux tubes (strings)  in supersymmetric QCD, such as $2D$-$4D$ 
	correspondence (see, {\it e.g.}, the review publications
	\cite{Trev,SYrev})
	attract exceeding attention now \cite{GGS}.
	A wide variety of non-perturbative effects was addressed in theories 
	which support such flux tubes
	\cite{Shifman:2010id}.
	Supersymmetry plays a special role in a number of aspects.
	Typically, the flux tubes require the existence of scalar fields. 
	\ntwon\, supersymmetric QCD supplies both scalar quarks and adjoint scalars. In addition,
	the power of supersymmetry manifests itself in providing a setting for obtaining
	 exact results (see, {\it e.g.}, Refs.
	\cite{Novikov:1983uc,Seiberg:1994rs,Seiberg:1994aj,Shifman:2013ewa}).

	The string becomes {\it non-Abelian} if it gives rise to the so-called orientational moduli living on
	its world sheet \cite{HT1,ABEKY,SYmon,HT2}. In the context of gauge theories this typically requires U$(N)_C\times$SU(N)$_F$
	spontaneously broken down to color-flavor locked  diagonal $\!$\cfl.
	Then the orientational moduli  span a \cpn space, and 
	the latter becomes the target space of the two--dimensional theory on the world sheet \cite{SYrev}.

	A soft breaking of \ntwon\, supersymmetry down to
	\nonen\, in the bulk gives rise to a richer  set of theories on the string world sheet.
	For the most part in this paper, however, we will deal with the \ntwon\, gauge theory.

	When non-vanishing (s)quark mass parameters are introduced in  the bulk theory, 
	the global \cfl group is explicitly broken, and, strictly speaking,
	the non-Abelian strings are no longer.
	The moduli parameters are lifted, and a shallow potential is generated. The only true minima  are the so-called $ \mc{Z}_N $ strings.
	In terms of the two--dimensional world sheet theory, these strings are described by
	the vacua of the two--dimensional potential.

	The fact of its existence and the form of this potential has been known for a long time 
\cite{Dorey:1998yh,Shifman:2006bs}.
	Indeed, the only form compatible with \ntwot supersymmetry in two dimensions is
\beq
\label{V1+1}
	V_{1+1}^\text{twisted-mass}    ~~=~~    \sum\, | m_j |^2\, \big| n_j \big|^2   ~-~  \Big| \sum\, m_j\, \big| n_j \big|^2 \,\Big|^2.
\eeq
	Here $ m_j $ are the mass parameters and $ n_j $ the orientational ({\it quasi}) moduli.
	On geometrical grounds, this potential was found in \cite{HT2} by Hanany and Tong. 
	Derivation of this potential from the bulk theory was only carried out in the SU(2) case \cite{SYmon}.
	As we will discuss below, the quark mass parameters induce a non-vanishing expectation value for the adjoint fields.
	An \ansatz was proposed for the adjoint field $ a^\text{SU(2)} $ in \cite{SYmon},
	the substitution of which into the bulk action produced the expected result \eqref{V1+1}.
	This paper extends the SU$(2)_C\times$SU(2)$_F$ bulk theory to the general case of SU$(N)_C\times$SU(N)$_F$.
	We propose an \ansatz for the adjoint fields in the general case  for the first time.
	We confirm our expressions by substituting the adjoints into the bulk action. 
	This procedure produces a consistent expression both for the two--dimensional action and for its normalization integral
	and in this way provides us with a direct
	derivation of the world-sheet potential \eqref{V1+1}.

	As another application of our \ansatz for the adjoint fields, we are able to confirm the potential 
\beq
\label{Vmu}
	V_{1+1}    \:~~ = ~~\:
	4\pi\, \Big|\, \muU\, m  \,~-~\,  \mu_N\, \big( \sqrt{2}\,\sigma \,+\, m \big) \,\Big|\,,
\eeq
	arising on the world sheet \cite{Shifman:2010kr} 
	once the \ntwon\, supersymmetry in the bulk is broken by a quadratic superpotential for the
	adjoint superfields $ \mu\, \ca^2 $ down to ${\mathcal N}=1$.
	Here $\sigma$ is a scalar field of the gauge multiplet in the world-sheet \cpn model, $m$ is the average quark mass,
	while $\mu_U$ and $\mu_N$ are the mass terms of the bulk U(1) and SU$(N)$ adjoint matter, respectively. 
	This potential becomes nontrivial once the quark masses are non-degenerate and breaks \ntwot world-sheet supersymmetry down to 
	\ntwoon. 
	
	For the case of a single-trace bulk deformation operator (\ie $\muU = \mu_N$) this 
	potential acquires a particularly simple form
	\beq
\label{Vmu1}
	V_{1+1}   ~~=~~
	4 \pi\, \Big|\,  \sqrt{2}\, \mu_N\, \sigma  \,\Big|\,.
\eeq
	Although our derivation is valid only to the linear order in $ \mu $, it is carried out
	starting directly from the bulk theory.

\section{Adjoint Fields}
\setcounter{equation}{0}

	We start with the \ntwo\, SQCD with $ N_f \,=\, N_c \,=\, N $ flavors transforming according to the  fundamental
	representation of the gauge group U(1)$\,\times\,$SU($N$).
	In order for the theory to support  non-Abelian strings, we introduce the Fayet-Illiopolous (FI) terms into the
	theory.
	The bosonic part of the Lagrangian is as follows,
\begin{align}
\notag
	\ml    & ~~=~~
	\frac{1}{2\, g_2^2}\, \Tr\, \Big( F_{\mu\nu}^\text{SU($N$)} \Big)^2  ~~+~~
	\frac{1}{g_1^2}\, \Big( F_{\mu\nu}^\text{U(1)} \Big)^2  ~~+~~
	\\[3mm]
\notag
	&
	~~+~~
	\frac{2}{g_2^2}\, \Tr\, \Big| \md_\mu\, \aN \Big|^2  ~~+~~
	\frac{4}{g_1^2}\, \Big| \p_\mu\, \aU \Big|^2  ~~+~~
	\\[3mm]
\label{bulk}
	&
	~~+~~
	\Tr\, \Big|\, \md_\mu\, q \,\Big|^2  ~~+~~
	\Tr\, \Big|\, \md_\mu\, \bqt \,\Big|^2
	~~+~~
	\\[4mm]
\notag
	&
	~~+~~
	V \big(\, q,\, \qt,\, \aU,\, \aN \,\big) \,.
\end{align}
	Here $ F_{\mu\nu}^\text{SU($N$)} $ and $ F_{\mu\nu}^\text{U(1)} $ are the field strengths
	of the non-Abelian and Abelian gauge fields correspondingly, 
	and $ \aN $ and $ \aU $ are the scalar adjoint fields (scalar superpartners of the gauge fields).
	The quark fields $ q $ and $ \qt $ which comprise the quark hypermultiplet are written in the
	color--flavor matrix notation (the first index of such a matrix refers to color and the second
	to flavor).
	The potential in the theory with \ntwo\, supersymmetry is
\begin{align}
\notag
	&
	V \big(\, q,\, \qt,\, \aU,\, \aN \,\big) ~~=~~ 
	\\[1mm]
\notag
	&
	~~=~~
	g_2^2\, \Tr \lgr \frac{1}{g_2^2}\, \Big[ \aN\: \baN \Big]  ~+~  \frac{1}{2}\, \Ts\, \left( q\,\bq \:-\: \bqt\,\qt \right) \rgr^2 
	~~+~~
	\\[1mm]
\notag
	&
	~~+~~
	\frac{g_1^2}{8} \lgr  \Tr \left( q\,\bq \:-\: \bqt\,\qt \right) ~-~ N\,\xi_3  \rgr^2
	~~+~~
	\\[2mm]
\label{pot}
	&
	~~+~~
	g_2^2\, \Tr\, \Big| \Ts\: q\,\qt \,\Big|^2
	~~+~~
	\frac{g_1^2}{2}\, \Big|\, \Tr\: q\,\qt  ~-~  \frac{N}{2}\,\xi \,\Big|^2
	~~+~~
	\\[2mm]
\notag
	&
	~~+~~
	2\, \Tr\, \bigg| \lgr \aU \:+\: \aN \rgr q  ~+~  q \cdot \frac{\mhat}{\sqrt{2}} \,\bigg|^2
	~~+~~
	\\[1mm]
\notag
	&
	~~+~~
	2\, \Tr\, \bigg| \lgr \aU \:+\: \aN \rgr \bqt  ~+~  \bqt \cdot \frac{\mhat}{\sqrt{2}} \,\bigg|^2
	\,.
\end{align}
	Here $ \Ts $ takes a traceless part of an expression.
	Parameter $ \xi_3 $ denotes the (real) $ D $-term FI parameter, while $ \xi $ is the (complex) $ F $-term FI parameter.
	When the \ntwo\, supersymmetry is not broken, these parameters are equivalent, and only one is necessary.
	We will therefore only use $ \xi_3 $, but will still call it $ \xi $ for brevity.
	Matrix $ \mhat $ here denotes the diagonal matrix of the quark mass parameters,
\beq
	\mhat    ~~=~~    \lgr\, \begin{matrix}
			          m_1  &       &        &       \\
				       &  m_2  &        &       \\
				       &       & \ddots &       \\
                                       &       &        &  m_N  
			         \end{matrix} \,\rgr
	.
\eeq
	Because this is a matrix in the flavor space, it multiplies matrix $ q $ on the right.
	For the theory to be accessible semiclassically, we canonically assume the FI parameter to be large,
\[
	\sqrt{\xi}    ~~\gg~~    \LN\,,~  m \,.
\]

\subsection{Zero masses}
	We start from the case in which the (s)quark masses vanish.
	Again, in this section we assume the FI $ F $-term equal to zero, with the $ D $-term denoted as
\[
	\xi_3    ~~\equiv~~    \xi    ~~\neq~~    0\,.
\]
	When the bare quark mass matrix vanishes,
\[
	\mhat    ~~=~~    0\,,
\]
	the theory supports non-Abelian string solutions.
	We will not   review  the perturbative spectrum of this model, referring the reader to \cite{SYrev}.
	We will just point out that the $ r \,=\, N $ vacuum of the potential \eqref{pot} can always be chosen
	in the color-flavor locked form
\beq
\label{qvac}
	\langle q^{kA} \rangle    ~~=~~    \sqrt{\xi}\, 
		\lgr \begin{matrix}
		     		   1    &    0     &    \ldots  \\
				\ldots  &  \ldots  &    \ldots  \\
				\ldots  &    0     &       1
		     \end{matrix} \rgr,
	\qquad\qquad
	\langle \qt{}_{Ak} \rangle    ~~=~~    0\,.
\eeq
	As currently we hold $ \mhat \,=\, 0 $, the adjoint fields vanish in this vacuum,
\beq
	\langle \aN \rangle    ~~=~~    \langle \aU \rangle    ~~=~~    0\,.
\eeq

	The string solutions are found as profile functions of the quark and gauge fields, which tend
	to the vacuum values at the infinity, but with a winding of one of their components 
	in the plane perpendicular to the string --- that is what keeps the string stable.
	The string \ansatz for the scalar fields is
\begin{align}
\notag
	q    & ~~=~~    \bq    ~~=~~    \phi\,,    \\[2mm]
\label{qansatz}
	\qt    & ~~=~~    \bqt    ~~=~~    0\,,    \\[2mm]
\notag
	\aU    & ~~=~~    \aN    ~~=~~    0\,.
\end{align}
	The quark matrix $ \phi $ is described in terms of the profile functions $ \phi_1(r) $ and $ \phi_2(r) $,
\beq
	\phi(r)    ~~=~~    \phi_2  ~~+~~  \nnbar \cdot (\, \phi_1 ~-~ \phi_2 \,)\,.
\eeq
	We chose here a singular gauge in which the quarks do not wind at all, but the gauge fields do,
	for which purpose they have to be singular at the core of the string $ r \,=\, 0 $.
	The \ansatz for the gauge fields is
\begin{align}
\notag
	A_j^\text{SU($N$)}    & ~~=~~    \epsilon_{jk}\, \frac{x^k}{r^2}\, f_N(r) \lgr \nnbar ~-~ 1/N \rgr ,
	\\
\label{Aansatz}
	A_j^\text{U(1)}    & ~~=~~    \frac{1}{N}\, \epsilon_{jk}\, \frac{x^k}{r^2}\, f(r)\,.
\end{align}
	These string profiles obey the first-order (BPS) equations
\begin{align}
\notag
	\p_r\, \phi_1(r)    & ~~=~~    \frac{1}{N\,r} \lgr f(r) ~+~ (N - 1)\,f(r) \rgr \phi_1(r)\,,
	\\[2mm]
\notag
	\p_r\, \phi_2(r)    & ~~=~~    \frac{1}{N\,r} \lgr f(r) ~-~ f_N(r) \rgr \phi_2(r)\,,
	\\[2mm]
\label{BPS}
	\p_r\, f(r)    & ~~=~~    \frac{N\, g_1^2}{4}\, r \lgr \phi_1(r)^2  ~+~ (N-1)\,\phi_2(r)^2  ~-~  N\,\xi \rgr ,
	\\[2mm]
\notag
	\p_r\, f_N(r)    & ~~=~~    \frac{g_2^2}{2}\, r \lgr \phi_1(r)^2  ~-~  \phi_2(r)^2 \rgr ,
\end{align}
	supplemented with the appropriate boundary conditions
\begin{align}
\notag
	  \phi_1(0)    & ~~=~~    0\,,
	& \phi_2(0)    & ~~\neq~~    0\,,
	& \phi_1(\infty)    & ~~=~~ \sqrt{\xi}\,,
	& \phi_2(\infty)    & ~~=~~ \sqrt{\xi}\,,
	\\[2mm]
	f_N(0)    & ~~=~~ 1\,,
	& f(0)    & ~~=~~ 1\,,
	& f_N(\infty)    & ~~=~~ 0\,,
	& f(\infty)    & ~~=~~ 0\,.
\end{align}
	The latter conditions at infinity ensure that the fields tend to their vacuum values, while
	the conditions in the string core are needed for the finiteness of the string tension 
	(and do not restrict the value of $ \phi_2(0) $ other than that it cannot vanish).

	The above \ansatz describes a family of solutions, labeled by the \cpn moduli variables $n^l$,
\beq
	\vec{n}    ~~\in~~    \mc{C}^N\,,
	\qquad\qquad
	\big|\, \vec{n} \,\big|^2    ~~=~~    1\,.
\eeq
	These so-called {\it orientational} moduli ``rotate'' the solution in the \sunu space.
	Each solution actually breaks the color-flavor group \cfl down to SU($N-1$)$\times$U(1).
	Thus, there are as many as
\beq
	\frac{\text{SU($N$)}}{\text{SU($N-1$)$\times$U(1)}}    ~~\sim~~    \text{CP$^{N-1}$}
\eeq
	solutions, which are labeled by the vector $ \vec{n} $.
	Note that in the \ansatz \eqref{qansatz}--\eqref{Aansatz}, in our notation, $ \nnbar $ is a matrix.

	It is these moduli, that give the string the name non-Abelian.
	They live on this string.
	In order to see this, one allows them to be weekly
	dependent on $ t $ and $ z $ (longitudinal) coordinates.
	Then it can be shown \cite{SYrev} that the bulk theory induces a ``live'' action for $ \vec{n} $
	on the world sheet of the strings.
	The way this happens is that when $ t $, $ z $ dependence is introduced, 
	the \ansatz \eqref{Aansatz} has to be extended --- 
	the longitudinal components of the gauge field now get excited,
\beq
\label{Amuansatz}
	A_\mu^\text{SU($N$)}    ~~=~~    i\, \big[\, \nnbar,\, \p_\mu(\nnbar) \,\big]\, \rho(r)\,,
	\qquad\qquad
	\mu ~=~ 0,\, 3\,.
\eeq
	Here $ \rho(r) $ is a new profile function with a boundary condition
\beq
	\rho(0)    ~~=~~    1\,,
\eeq
	which again is needed for finiteness of the string tension.
	When now all the profiles \eqref{qansatz}--\eqref{Aansatz} and \eqref{Amuansatz} are substituted into the
	bulk action \eqref{bulk}, and integrated over the transverse coordinates, the following theory emerges on the
	world sheet of the string,
\beq
\label{ws}
	S    ~~=~~    2\,\beta\, \int\, d^2x\, \lgr \big|\, \p_\mu\, n \,\big|^2  
					~+~  \big( \nbar\, \p_\mu\, n \big)^2 \rgr,
\eeq
	with the summation index $ \mu $ running over the longitudinal coordinates (0 and 3).
	Here $ \beta $ is a normalization constant, arising due to the transverse integration of the profile functions,
\begin{align}
\label{betanorm}
	&
	\qquad
	\beta    ~~=~~    	\frac{2\pi}{g_2^2}\, \times
	\\
\notag
	&
	\int r\, dr
		\Biggl\lgroup\!  (\p_r\, \rho)^2 \,+\, \frac{1}{r^2}\, f_N^2\, (1-\rho)^2 
			~+~  g_2^2 \lgr (1-\rho)\, \big( \phi_1 \,-\, \phi_2 \big)^2  ~+~
				\frac{1}{2}\, \rho^2\, \big(\, \phi_1^2 \,+\, \phi_2^2 \,) \rgr \!
		\!\Biggr\rgroup,
\end{align}
	and effectively becoming the coupling constant of the two-dimensional theory.
	Minimization of \eqref{betanorm} with respect to $ \rho(r) $ gives
\beq
	\rho(r)    ~~=~~    1  ~~-~~ \frac{\phi_1}{\phi_2}\,.
\eeq
	If one now takes into account the BPS equations \eqref{BPS} for the profiles, then the 
	second line in Eq.~\eqref{betanorm} reduces to unity, and
\beq
	\beta    ~~=~~    \frac{2\pi}{g_2^2}\,.
\eeq
	
	Note that the action \eqref{ws} could and would actually have higher--order derivative corrections, running in powers of
\beq
\label{higher}
	\frac{\p_\mu}{g_2\sqrt{\xi}}\,.
\eeq
	Below the scale of the inverse thickness of the string, $ g_2\,\sqrt{\xi} $, where the world-sheet
	description \eqref{ws} is valid, such corrections are negligible.

\subsection{Nonvanishing masses}

	When non-zero masses are introduced in the theory \eqref{bulk}, the situation changes significantly.
	The non-Abelian strings cease to be solutions of equations of motion, 
	and the orientational moduli $ \vec{n} $ are lifted\footnote{These moduli are lifted at the quantum level even if all mass terms vanish. 
	But this is a quantum effect.
	The above statement can be reformulated more accurately as follows: 
	the orientational moduli $ \vec{n} $ are lifted at the classical level if $m_i\neq m_j\neq 0$.}.
	They become {\it quasi}--moduli, as a shallow potential is generated on the world sheet.
	Only when $ \vec{n} $ equals one of 
\beq
\label{nvac}
	\nvac    ~~=~~    (\, 0\,,~ \dots\,,~1\,,~ \dots\,,~ 0 \,)
\eeq
	does the string become a BPS solution again, in the sense of the low energy Abelian theory.
	As there are $ N $ such strings, they are called the $ \mc{Z}_N $ strings.

	The \ansatz for the squarks and gauge fields remains the same
\begin{align}
\notag
	q    & ~~=~~    \bq    ~~=~~    \phi\,,    \\[4mm]
\notag
	\qt    & ~~=~~    \bqt    ~~=~~    0\,,    \\[2mm]
\label{qAansatz}
	A_j^\text{SU($N$)}    & ~~=~~    \epsilon_{jk}\, \frac{x^k}{r^2}\, f_N(r) \lgr \nnbar ~-~ 1/N \rgr ,
	\\[2mm]
\notag
	A_j^\text{U(1)}    & ~~=~~    \frac{1}{N}\, \epsilon_{jk}\, \frac{x^k}{r^2}\, f(r)\,.
\end{align}
	The first obvious change, revealed by inspecting the last two lines of Eq.~\eqref{pot}, 
\begin{align}
\notag
	&
	\phantom{~~+~~}
	2\, \Tr\, \bigg| \lgr \aU \:+\: \aN \rgr q  ~+~  q \cdot \frac{\mhat}{\sqrt{2}} \,\bigg|^2
	~~+~~
	\\[1mm]
\label{Fterms}
	&
	~~+~~
	2\, \Tr\, \bigg| \lgr \aU \:+\: \aN \rgr \bqt  ~+~  \bqt \cdot \frac{\mhat}{\sqrt{2}} \,\bigg|^2
	\,,
\end{align}
	is that the vacuum values of the adjoint scalars are no longer zero,
\begin{align}
\notag
	\aU    ~~=~~    \langle \aU \rangle    & ~~=~~    -\, \frac{m}{\sqrt{2}}\,,
	\\
\label{avac}
	\langle \aN \rangle    & ~~=~~ -\, \frac{\dm}{\sqrt{2}}\,,
\end{align}
	Here $ m $ is the average mass parameter, and $ \dm $ is the diagonal matrix of the mass differences,
\beq
	\dm{}_j    ~~=~~    \mhat{}_j  ~~-~~  m\,,
	\qquad\qquad
	m    ~~=~~ \frac{1}{N}\, \sum\, \mhat{}_j\,.
\eeq
	
	The second ``massive'' $ F $-term in Eq.~\eqref{Fterms} is responsible 
	for making the non-Abelian string a {\it quasi}-solution, 
	except when $ \vec{n} $ takes one of its vacuum values \eqref{nvac}.
	
	As is shown in the first line of Eq.~\eqref{avac}, the U(1) scalar $ \aU $ does not develop
	any profile and always sits in its vacuum.
	Its sole purpose is to cancel the average mass $ m $ in the above $ F $-terms 
	(since the average mass is essentially a unit matrix, it commutes with $ q $, and the 
	cancellation occurs everywhere). In fact, the average quark mass can be 
	eliminated by the shift of $ \aU $.

	A very different thing happens to the SU($N$) field $ \aN $.
	As the average mass has been canceled everywhere, it is only $ \dm $ that is left to cancel.
	However, the latter is not generically proportional to the unit matrix, and so the complete cancellation
	can only happen at infinity (or whenever $ \vec{n} ~=~ \nvac $, in which
	case $ q $ commutes with everything).
	Therefore field $ \aN $ does have a profile, 
	which asymptotically tends to the vacuum value given by the mass differences in Eq.~\eqref{avac}.

	The \ansatz for the non-Abelian adjoint field $ \aN $ has been known for the case
	of the SU(2) gauge group \cite{SYmon}.
	In this case the CP$^1$ moduli variables $ n^l $ can be traded for O(3) variables $ S^a $,
\beq
\label{S}
	S^a    ~~=~~    (\, \nbar\: \tau^a\, n \,)\,.
\eeq
	In terms of these, the known \ansatz looks as
\beq
\label{su(2)}
	a^\text{SU(2)}    ~~=~~
	a^a\, \frac{\tau^a}{2}    ~~=~~    -\, \frac{\deltam}{\sqrt{2}}\, 
		\lgr  \tau^3\, \omega(r)  ~+~  S^3\, S^a\,\tau^a\, (1 \,-\, \omega(r))  \rgr.
\eeq
	Here $ \deltam $ is the only mass difference $ (m_1 ~-~ m_2)/2 $, and the reason
	that the third direction enters explicitly is because $ \dm ~\propto~ \tau^3 $ in this case.
	The profile function $ \omega(r) $ satisfies the following boundary conditions
\beq
	\omega(0)    ~~=~~    0\,,
	\qquad\qquad
	\omega(\infty)    ~~=~~    1\,,
\eeq
	and is found by a minimization procedure, giving
\beq
	\omega(r)    ~~=~~    \frac{\phi_1(r)}{\phi_2(r)}\,.
\eeq
	The r\^ole of this profile function is to give $ a^\text{SU(2)} $ an interpolation
	between the vacuum value (when $ \omega ~=~ 1 $),
\beq
	a^\text{SU(2)}(\infty)    ~~=~~    -\, \frac{\dm}{\sqrt{2}}    ~~=~~    -\, \frac{\deltam \cdot \tau^3}{\sqrt{2}}
\eeq
	and its value at the core of the string (when $ \omega ~=~ 0 $),
\beq
	a^\text{SU(2)}(0)    ~~=~~    -\, \frac{\deltam}{\sqrt{2}}\, S^3\, \big( S^a\, \tau^a \big)\,.
\eeq
	The latter expression is proportional to $ S^a\, \tau^a $, and commutes with the gauge field
	(which is proportional to the same matrix structure).
	This is needed so that the kinetic term of $ a^\text{SU(2)} $ containing the
	commutator $ \Big[ A^\text{SU(2)}_\mu \,,\, a^\text{SU(2)} \Big] $ does not produce 
	a divergent contribution to the string tension, due to the singularity of the gauge field at the core.
	At the same time, if $ \vec{S} $ happens to be parallel to the third axis (\ie the string is in the vacuum),
	then $ \omega(r) $ in Eq.~\eqref{su(2)} cancels away, and the adjoint field takes its vacuum value everywhere in the space.

	We now give the generalization of the \ansatz \eqref{su(2)} to the case
	of the SU($N$) gauge group.
	The expression appears to be more involved than its SU(2) counterpart, namely,
\beq
\label{ansatz}
	\aN    ~~=~~    
	-\, \frac{1}{\sqrt{2}}\, 
	\lgr \dm  ~-~  (1 \,-\, \omega(r)) \Big[\, \nnbar\, \big[\, \nnbar,\, \dm \,\big] \,\Big] \rgr.
\eeq
	We will show that $ \omega(r) $ is the same profile function as in Eq.~\eqref{su(2)}.

	Before discussing the properties of this \ansatz we first bring a few useful relations
	involving matrix $ \nnbar $.
	These relations owe to the fact that 
\beq
\label{nnbarsq}
	\big( \nnbar \big)^2 ~~=~~ \nnbar\,.
\eeq
	We notice that expression \eqref{ansatz} involves the second commutator of $ \nnbar $
	and the mass difference matrix $ \dm $.
	It appears that the \emph{third} commutator of $ \nnbar $ and any matrix actually
	equals the first commutator of these,
\beq
\label{comm}
	\Big[\, \nnbar\, \Big[\, \nnbar\, \big[\, \nnbar,\, \hat{M} \,\big] \,\Big] \,\Big]    ~~=~~
	\big[\, \nnbar,\, \hat{M} \,\big]\,.
\eeq

	Expression \eqref{ansatz} takes the vacuum value $ \dm $ at infinity, and 
	``rotates'' it as $ r $ goes to zero.
	The only available ``color'' parameter for such a rotation is $ \nnbar $.
	Let us show that indeed such a rotation takes place.
	Note that, because of the property \eqref{nnbarsq}, an exponent involving $ \nnbar $
	will always reduce to trigonometric functions.
	Then a ``rotation'' of any matrix $ \hat{M} $ will look as follows,
\beq
	e^{i \alpha \nnbar} \cdot \hat{M} \cdot e^{- i \alpha \nnbar}    ~~=~~
	\hat{M}  ~~+~~  i\, \sin \alpha \: \big[\, \nnbar,\, \hat{M} \,\big]
	~~-~~  ( 1 ~-~ \cos \alpha )\, \Big[\, \nnbar\, \big[\, \nnbar,\, \hat{M} \,\big] \,\Big].
\eeq
	Getting rid of the imaginary part, expression \eqref{ansatz} can then be written as
\beq
\label{rotation}
	-\, \sqrt{2} \cdot \aN    ~~=~~    
	\frac{1}{2}\, e^{i \alpha \nnbar} \cdot \dm \cdot e^{- i \alpha \nnbar}  ~~+~~
	\frac{1}{2}\, e^{- i \alpha \nnbar} \cdot \dm \cdot e^{i \alpha \nnbar}\,,
\eeq
	where 
\beq
	\cos\alpha(r)    ~~=~~    \omega(r)\,.
\eeq
	Another way of writing this is to notice that an exponent of commutators of $ \nnbar $
	with any matrix (\ie a commutator exponent analogous to that in the kinetic term
	of the adjoint scalar) will similarly be reducible to trigonometric functions owing
	to Eq.~\eqref{comm}. 
	Then our \ansatz can be written as a ``cosine''
\beq
	-\, \sqrt{2} \cdot \aN    ~~=~~
	\frac{ e^{ i\, \alpha\, [\, \nnbar \; \cdot \;] } 
	   ~+~ e^{ - i\, \alpha\, [\, \nnbar \; \cdot \;] } } { 2 }\, \dm\,.
\eeq

	Now let us discuss the properties of this {\it ansatz}. 
	First of all, it is easy to see that it is a traceless matrix.
	Next, we repeat, as $ r $ goes to infinity ($ \omega(r) ~\to~ 1$, and $ \alpha ~\to~ 0 $),
	the adjoint field approaches the vacuum value
\beq
	\aN    ~~\overset{r\,\to\,\infty}{\longrightarrow}~~    
	\langle \aN \rangle    \;~~=~~\;    -\, \frac{\dm}{\sqrt{2}}\,.
\eeq
	On the other hand, at the core of the string, the solution turns into a matrix
\beq
	-\, \sqrt{2} \cdot \aN(0)    ~~=~~    \dm  ~~-~~  \Big[\, \nnbar\, \big[\, \nnbar,\, \dm \,\big] \,\Big]\,,
\eeq
	which, because of property \eqref{comm}, commutes with $ \nnbar $.
	This way,  at the string core the adjoint field commutes with the gauge field 
	(proportional to $ \nnbar ~-~ 1/N $, see Eq.~\eqref{qAansatz}), 
	and the gauge field singularity is avoided.
	Note that, unlike in the case of SU(2), the adjoint field does not become proportional solely to $ \nnbar ~-~ 1/N $
	at the core.

	It is also easy to check the BPS condition on the solution \eqref{ansatz}.
	Indeed, when $ \vec{n} ~=~ \nvac $, matrix $ \nnbar $ commutes with anything,
	and the right hand side in Eq.~\eqref{ansatz} reduces to the vacuum value
\beq
	\aN (\nvac)    ~~=~~    \langle \aN \rangle    ~~=~~    -\, \frac{\dm}{\sqrt{2}}
\eeq
	everywhere in the space.

	Finally, it is slightly more technical, but straightforward, to check that Eq.~\eqref{ansatz}
	reduces to Eq.~\eqref{su(2)} for the gauge group SU(2), {\it i.e.}, is a correct generalization.

	The \ansatz \eqref{ansatz} is not the only generalization of the SU(2) formula \eqref{su(2)}.
	In fact, if one took the ``direct'' correspondence rules (see the definition \eqref{S})
\begin{align}
\notag
	\deltam\, \tau^3    & ~~\longrightarrow~~    \dm\,,\\[2mm]
	\frac{S^a\, \tau^a}{2}    & ~~\longrightarrow~~    \nnbar \,-\, 1/N \,,\\[3mm]
\notag
	S^3    & ~~\longrightarrow~~     (\nbar\, \tau^3\, n)\,,
\end{align}
	and applied them to \eqref{su(2)}, the following expression would emerge,
\[
	-\, \frac{1}{\sqrt{2}}\, \lgr  \dm \cdot \omega(r)  \;~+~\;  
	2\, ( 1 - \omega(r) ) \cdot (\nbar\, \dm\, n) \, \big( \nnbar \,-\, 1/N \big) \rgr.
\]
	The latter expression certainly does reduce to \eqref{su(2)} if one  again assumes $ N = 2 $.
	However, this expression does not work for generic $N$.
	Most obvious is the fact that it does not satisfy the BPS condition ---
	it does not reduce to the constant vacuum value when $ \vec{n} ~=~ \nvac $.

	At the same time, when  one takes \eqref{ansatz} and substitutes it into the bulk action \eqref{bulk},
	the following potential emerges on the world sheet of the string:
\begin{align}
\notag
	&
	\frac{4\pi}{g_2^2}\, \int\, r\, dr
		\Biggl\lgroup  (\p_r\, \omega)^2 \,+\, \frac{1}{r^2}\, f_N^2\, \omega^2 
			~+~  g_2^2 \lgr \omega\, \big( \phi_1 \,-\, \phi_2 \big)^2  ~+~
				\frac{1}{2}\, (1 - \omega)^2\, \big(\, \phi_1^2 \,+\, \phi_2^2 \,) \rgr
		\!\Biggr\rgroup
	\\
\label{asub}
	&
	\qquad\quad~
	\times\,
	\int\, d^2x\,
	\lgr  \big( \nbar\, \big| \dm \big|^2\, n \big)  ~-~  \big|\, ( \nbar\, \dm\, n ) \,\big|^2  \rgr
	~~+~~
	O \big( \dm^4 \big).
\end{align}
	We notice that the normalization integral here appears to be the same as in Eq.~\eqref{betanorm},
	which, therefore, gives us via minimization, 
\beq
\label{omega}
	\omega(r)    ~~=~~    1  ~~-~~ \rho(r)    ~~=~~    \frac{\phi_1(r)}{\phi_2(r)}\,,
\eeq
	and the whole integral in the first line of expression \eqref{asub} reduces to unity.
	As for the corrections $ O \big( \dm^4 \big) $, they look as (here 
	the representation \eqref{rotation} is helpful in finding their form)
\beq
	O \big( \dm^4 \big)    ~~=~~    
	2\pi\, \int\, r\, dr\, \frac{1}{2}\, \big( 1 \,-\, \omega^2 \big)^2 \,\cdot\, \big( \dm \big)^4 \,,
\eeq
	where $ \big( \dm \big)^4 $ is an expression involving $ \vec{n} $ and the fourth power of $ \dm $.
	The profile integral in the above expression is saturated at the thickness of the string.
	Therefore, by dimensional counting, these corrections are suppressed by a power of $ \xi $,
\beq
	O \big( \dm^4 \big)    ~~\sim~~    \big| \dm \big|^2 \cdot \frac{ \big| \dm \big|^2 }{ g_2^2\, \xi }\,,
\eeq
	and can be ignored on the same grounds as the higher--order derivatives \eqref{higher}.

	Taking Eq.\eqref{omega} into account, we write the result for $ \aN $ as
\beq
\label{aN}
	\aN    ~~=~~    
	-\, \frac{1}{\sqrt{2}}\, 
	\lgr \dm  ~-~  \rho(r)\, \Big[\, \nnbar\, \big[\, \nnbar,\, \dm \,\big] \,\Big] \rgr.
\eeq
	We observe that this expression provides us with the expected form of the 
	twisted--mass potential on the world sheet of the string,
\begin{align}
\notag
	& 2\beta\, \int\, d^2x \lgr\,  \big( \nbar\, \big| \dm \big|^2\, n \big)  ~-~  \big|\, ( \nbar\, \dm\, n ) \,\big|^2  \,\rgr
	~~=~~ 
	\\
\label{twpot}
	~~=~~ &
	2\beta\, \int\, d^2x \lgr\, \sum\, | m_k |^2\, \big| n_k \big|^2   ~-~  \Big|\, \sum\, m_k\, \big| n_k \big|^2 \,\Big|^2 \,\rgr.
\end{align}
	Here we use  the well--known shift invariance of this potential in order to replace $ \dm{}_k $ by $ m_k $.

	To conclude this section we note that the twisted--mass--deformed \cpn model can be nicely rewritten as a
	strong coupling limit of a U(1) gauge theory \cite{HaHo}. 
	In this description the meaning of the twisted--mass potential becomes  transparent. 
	Namely, the potential  reduces to the mass terms for $\vec{n}$-fields.  
	The bosonic part of the action reads
\beqn
	S 
	& ~=~ &
	\int d^2 x \left\{\,
	 2\beta\,|\nabla_{\mu} n_{k}|^2 ~+~ \frac1{4e^2}\, F^2_{\mu\nu} ~+~ \frac1{e^2}\, |\pt_{\mu}\sigma|^2
	\right.
\nonumber\\[3mm]
	&& \quad~ 
	~+~ \left. 2\beta\,| \sqrt{2}\sigma + m_k |^2\, |n_{k}|^2 ~+~
	2\, e^2 \beta^2\, (|n_{k}|^2 -1)^2
	\,\right\}\,.
\label{gaugecp}
\eeqn
	Here $\sigma$ is a scalar superpartner of the U(1) gauge field.
	In the limit $ e^2 \,\to\, \infty$, fields $A_{\mu}$ and $\sigma$ can be excluded by virtue of the algebraic
	equations of motion, namely
\beq
	A_{\mu} ~~=~~ -\, \frac{i}{2}\,\left(\, \nbar\, \pt_{\mu}n \,-\, \pt_{\mu}\nbar \,n \,\right),
\qquad 
	\sigma    ~~=~~    -\, \sum\, \frac{m_j}{\sqrt{2}}\, \big|\, n_j \,\big|^2\,.
\label{Asigma}
\eeq

	Substitution of this into (\ref{gaugecp}) brings us back to the \cpn model with the potential (\ref{twpot}).

\section{Potential on the Heterotic Vortex String}
\setcounter{equation}{0}

	One interesting kind of deformation of the \ntwon\, theory supporting vortex strings is
	achieved by introducing quadratic terms for the adjoint fields in the superpotential,
\beq
\label{WA}
	\mw{}_\ma    ~~\:\supset\:~~    \Tr \lgr\, \muU\, \big( \AU \big)^2   ~~+~~  \mu_N\, \big( \AN \big)^2 \,\rgr.
\eeq
	Here we have introduced parameters $ \muU $, $ \mu_N $ which are related to $ \mu_1 $, $ \mu_2 $ of \cite{SY1} via\footnote{
	One of the advantages of the new notation is that the so-called ``single--trace'' operator
	corresponds to the case $ \muU \,/\, \mu_N  ~=~ 1 $.
	We, however, will duplicate the key results in both notations.}
\beq
	\muU    ~~=~~    \sqrt{\frac{2}{N}}\, \mu_1\,,
	\qquad\qquad
	\mu_N    ~~\equiv~~    \mu_2\,.
\eeq
	Such a superpotential breaks supersymmetry to \nonen. 
	The world sheet theory on the heterotic vortex string was studied in detail in \cite{Edalati,SY1,BSY3} 
	for the bulk theory with massless quarks and non-zero FI $D$-term $ \xi_3 $ and in \cite{Shifman:2010kr} for the 
	theory with massive quarks and zero $ \xi_3 $.

	We have a chance now to directly confirm the moduli potential arising on string to the linear order
	in the supersymmetry--breaking parameters $ \muU $ and $ \mu_N $ \cite{Shifman:2010kr}.
	In such a theory, the FI $ F $-terms are induced implicitly, due to the superpotential \eqref{WA},
\beq
\label{FIFterms}
	\frac{1}{2}\,g_1^2~ \Big|\, \Tr~ q\, \qt ~+~ \sqrt{2}\, N\, \muU \cdot \aU \Big|^2
	~~+~~
	g_2^2~ \Tr\, \Big|\, \Ts~ q\, \qt ~+~ \sqrt{2}\, \mu_N \cdot \aN \,\Big|^2\,.
\eeq

	From now on we assume that $\xi_3=0$, while the effective FI $F$-components  $\xi$ are generated due to nonzero
	vacuum values \eqref{avac} of the adjoint fields. 
	In particular, the average quark mass cannot be excluded any longer. 
	It becomes a new parameter which determines the average quark condensate. 
	More precisely, classically the quark vacuum expectation values (VEVs) are determined by
\beq
\label{xij}
	\xi_j    ~~\approx~~    2\, \left(\, \muU\,m  ~~+~~  \mu_N\,\dm{}_j \,\right)\,.
\eeq

	If the quark mass differences vanish these parameters reduce to a single FI term 
	which does not break \ntwon\, supersymmetry in the bulk and 
	\ntwot supersymmetry on the world sheet in the linear order in $\mu$ \cite{HSZ,VY}. 
	However, once the quark mass differences are small but nonvanishing, the color--flavor group SU$_{C+F}(N)$ is broken 
	because both the adjoint and quark VEVs are no longer equal (\ie flavour--universal). 
	In this case a shallow potential is generated in the world--sheet \cpn model
	breaking \ntwot supersymmetry down to \ntwoo \cite{Shifman:2010kr}\footnote{Note, that this does not happen 
	in the theory with the FI $D$-term. 
	Namely, the twisted--mass potential of the previous section 
	does not break \ntwot supersymmetry on the world sheet.}. 
	The non--Abelian string becomes a heterotic string \cite{Edalati,SY1}.

	To derive the world--sheet potential, we substitute the expression \eqref{aN} into the $ F $-terms
	\eqref{FIFterms} and expand the latter to the linear order in $ \dm $.
	The first term in Eq.~\eqref{FIFterms} does not contain $ \dm $, 
	and is just part of the average (\ie zero--order) string tension
\beq
\label{avtension}
	2 \pi\, \big|\, \hat{\xi} \,\big|    ~~=~~    2 \pi \cdot \Big|\, 2\, \muU\, m \,\Big| \,.
\eeq
	
	As for the second term, we notice that when plugging in the adjoint field 
\[
	\aN    ~~=~~    
	-\, \frac{1}{\sqrt{2}}\, 
	\lgr \dm  ~-~  \rho(r)\, \Big[\, \nnbar\, \big[\, \nnbar,\, \dm \,\big] \,\Big] \rgr,
\]
	its commutator part does not contribute at the linear order --- 
	the traceless part of $ q\, \qt $ is proportional to $ \nnbar \,-\, 1/N $, and
\[
	\Tr~\, \nnbar \, \big[\, \nnbar \,,\, * \,\big]    ~~=~~    0\,.
\]
	Therefore, only the vacuum value $ \langle \aN \rangle $ plays a r\^ole here.
	The profile integral involving $ \phi_1(r) $ and $ \phi_2(r) $ in $ q\, \qt $
	reduces to an integral of a total derivative due to the BPS equations \eqref{BPS},
\[
	2\pi\, \int\, r\, dr\, g_2^2\, \big( \phi_1 \,-\, \phi_2 \big)^2    ~~=~~
	4\pi\, \int\, dr\, \p_r\, f_N(r)    ~~=~~    -\, 4\pi\,,
\] 
	and the resulting linear terms are   
\beq
	2\pi \cdot 
	\lgr
		\mu_N\, \big(\nbar\, \dm\, n\big) \cdot \frac{\ov{\muU\, m}}{\big|\, \muU\, m \,|}
		~+~
		\ov{\mu}{}_N\, \big(\nbar\, \dmdag\, n\big) \cdot \frac{\muU\, m}{\big|\, \muU\, m \,|}
	\rgr.
\eeq

	Now it is obvious that this expression comprises the linear terms in the expansion of
	the absolute value in a series in $ \dm $,
\begin{align}
	&
	V_{1+1}    ~~=~~    4 \pi \cdot \big|\, \muU\, m  ~+~  \mu_N\, \big(\nbar\, \dm\, n\big) \,\big|    ~~=~~    
	\\
\notag
	&
	\quad~
	~~=~~
	4 \pi \cdot 
	\lgr
		\muU\, m  
		~+~
		\mu_N\, \big(\nbar\, \dm\, n\big) \cdot \frac{\ov{\muU\, m}}{\big|\, \muU\, m \,|}
		~+~
		\ov{\mu}{}_N\, \big(\nbar\, \dmdag\, n\big) \cdot \frac{\muU\, m}{\big|\, \muU\, m \,|}
		~+~
		\dots
	\rgr\!.
\end{align}

	In terms of parameters $ \mu_1 $ and $ \mu_2 $, this formula reads,
\begin{align}
	&
	V_{1+1}    ~~=~~    4 \pi \cdot \bigg|\, \sqrt{\frac{2}{N}}\,\mu_1\, m  ~+~  \mu_2\, \big(\nbar\, \dm\, n\big) \,\bigg|    ~~=~~    
	\\
\notag
	&
	~~=~~
	4 \pi \cdot 
	\lgr
		\sqrt{\frac{2}{N}}\,\mu_1\, m  
		~+~
		\mu_2\, \big(\nbar\, \dm\, n\big) \cdot \frac{\ov{\mu_1\, m}}{\big|\, \mu_1\, m \,|}
		~+~
		\ov{\mu}{}_2\, \big(\nbar\, \dmdag\, n\big) \cdot \frac{\mu_1\, m}{\big|\, \mu_1\, m \,|}
		~+~
		\dots
	\rgr\!\!.
\end{align}

	The above formulas perfectly agree with the two--dimensional potential found in \cite{Shifman:2010kr}.
	Adding and subtracting $ \mu_N\, m ~=~ \mu_N\, m\, (\nbar\, n) $ inside the absolute value,
	and trading variables $ \vec{n} $ for an auxiliary variable $ \sigma $ via (\ref{Asigma})\footnote{Here we still can use Eq.~(\ref{Asigma})
	assuming that the $\mu$--induced potential $V_{1+1}$ is a small correction to the action (\ref{gaugecp}).}
	we have,
\begin{align}
	\notag
	V_{1+1}\, (\sigma)    & ~~=~~    4 \pi \cdot \Big|\, \muU\, m  \,~-~\,  \mu_N\, \big( \sqrt{2}\,\sigma \,+\, m \big) \,\Big|    ~~=~~    
	\\
\label{Vdeform}
	& ~~=~~
	4 \pi \cdot \Big|\, \sqrt{\frac{2}{N}}\,\mu_1\, m  \,~-~\,  \mu_2\, \big( \sqrt{2}\,\sigma \,+\, m \big) \,\Big|\,.
\end{align}

	Note that now (in contrast to the case of the FI $D$-term)
	the vacuum energies of this world sheet potential give  the string tensions,
\beq
 	T_j    ~~=~~    V_{1+1}(\sigma_j),
\label{tv}
\eeq
	where $\sigma_j $ are VEVs of the field $  \sigma $ in the $ N $ vacua of the \cpn model. 
	Classically $ \sqrt{2}\sigma_j \,=\, -m_j $. 
	This is the way the potential (\ref{Vdeform}) was conjectured in \cite{Shifman:2010kr}.
	Indeed, using Eq.~\eqref{xij} we find correct string tensions
\beq
	T_j    ~~=~~    2\pi\, |\xi_j |\,.
\eeq
	We can see that in the \cpn model with potential (\ref{Vdeform}) for generic quark masses 
	the \ntwoo supersymmetry of the action is broken by the choice of the vacuum already at the classical level. 
	The vacuum energies in the $ N $ vacua of the \cpn model are generically all different.

	To conclude this section, let us note that the potential 
	(\ref{Vdeform}) gives quantum corrections to the string tensions
	\cite{Shifman:2010kr}.
	In the quantum theory the VEV of the  $\sigma$ field in each of the $ N $ vacua of the \cpn
	model with a weak deformation (\ref{Vdeform}) is given by
	solutions of the equation \cite{AdDVecSal,ChVa,W93,HaHo}
\beq
	\prod_{i=1}^{N}(\sqrt{2}\sigma +m_i)=\Lambda_{CP}^N,
\eeq
	where $ \Lambda_{CP} $ is the scale of the \cpn model. Solutions $  \sigma_i$ to this equation  give exact string tensions
	via Eq.~(\ref{tv}) with all corrections 
	in powers of $ \Lambda_{CP} / m_i $ included.

\section{Conclusions}

	We found an expression (Eq.~\eqref{aN}) for the adjoint field profiles 
	for the non-Abelian vortex configuration in \ntwon\, supersymmetric QCD 
	with the gauge group U($N)$ and  $ N $ flavors.
	This expression  enabled us to derive the twisted--mass potential \eqref{twpot} on the 
	vortex world sheet starting from the bulk theory.

	In the case in which \ntwon\, supersymmetry is softly broken by an operator $ \mu\, \ma^2 $, 
	which at the same time stabilizes the string acting as an effective FI $ F $-term,
	we managed to use expression \eqref{aN} to derive and confirm to the linear order
	the potential \eqref{Vdeform} generated on the world sheet.
	Our result is in agreement with the potential found in \cite{Shifman:2010kr}
	and removes the ambiguity of adding a potential vanishing in the critical points of Eq.~\eqref{Vdeform}.

\section*{Acknowledgments}

	This work  is supported in part by DOE grant DE-FG02-94ER40823. 
	The work of A.Y. was  supported 
	by  FTPI, University of Minnesota, 
	by RFBR Grant No. 13-02-00042a 
	and by Russian State Grant for 
	Scientific Schools RSGSS-657512010.2.


\small
\begin{thebibliography}{99}
\itemsep -2pt

 \bibitem{Trev}
  D.~Tong,
  Annals Phys.\  {\bf 324}, 30 (2009)
  [arXiv:0809.5060 [hep-th]];
  M.~Eto, Y.~Isozumi, M.~Nitta, K.~Ohashi and N.~Sakai,
  J.\ Phys.\ A  {\bf 39}, R315 (2006)
  [arXiv:hep-th/0602170];
  K.~Konishi,
  Lect.\ Notes Phys.\  {\bf 737}, 471 (2008)
  [arXiv:hep-th/0702102];
  M. Eto, Y.~Hirono, M.  Nitta, and S. Yasui,
  {\em Vortices and Other Topological Solitons in Dense Quark Matter},
  arXiv:1308.1535 [hep-ph] (to be published in Prog. Theor. Phys.).

  \bibitem{SYrev}
  M.~Shifman and A.~Yung,
  {\em Supersymmetric Solitons,}
  (Cambridge University Press, 2009).
  
  \bibitem{GGS}
  D.~Gaiotto, S.~Gukov and N.~Seiberg,
  {\em Surface Defects and Resolvents,}
  arXiv:1307.2578 [hep-th].

  \bibitem{Shifman:2010id} 
  M.~Shifman and A.~Yung,
  Phys.\ Rev.\ D {\bf 81}, 085009 (2010)
  [arXiv:1002.0322 [hep-th]],
  P.~Koroteev, M.~Shifman, W.~Vinci and A.~Yung,
  Phys.\ Rev.\ D {\bf 84}, 065018 (2011)
  [arXiv:1107.3779 [hep-th]],
%
  M.~Eto, T.~Fujimori, S.~B.~Gudnason, Y.~Jiang, K.~Konishi, M.~Nitta and K.~Ohashi,
  JHEP {\bf 1112}, 017 (2011)
  [arXiv:1108.6124 [hep-th]],
%
  S.~Chen, R.~Zhang and M.~Zhu,
  arXiv:1201.1602 [math-ph],
%
  K.~Konishi, M.~Nitta and W.~Vinci,
  JHEP {\bf 1209}, 014 (2012)
  [arXiv:1206.4546 [hep-th]],
%
  K.~Konishi,
  arXiv:1209.1376 [hep-ph],
%
  M.~Eto, T.~Fujimori, M.~Nitta and K.~Ohashi,
  JHEP {\bf 1307}, 034 (2013)
  [arXiv:1207.5143 [hep-th]],
%
  Q.~-H.~Huo, Y.~Jiang, R.~-Z.~Wang and H.~Yan,
  Europhys.\ Lett.\  {\bf 101}, 27001 (2013)
  [arXiv:1202.4511 [cond-mat.supr-con]],
%
  S.~Yasui, Y.~Hirono, K.~Itakura and M.~Nitta,
  Phys.\ Rev.\ E {\bf 87}, 052142 (2013)
  [arXiv:1204.1164 [cond-mat.supr-con]].

  \bibitem{Novikov:1983uc} 
  V.~A.~Novikov, M.~A.~Shifman, A.~I.~Vainshtein and V.~I.~Zakharov,
  Nucl.\ Phys.\ B {\bf 229}, 381 (1983).

  \bibitem{Seiberg:1994rs} 
  N.~Seiberg and E.~Witten,
  Nucl.\ Phys.\ B {\bf 426}, 19 (1994)
  [Erratum-ibid.\ B {\bf 430}, 485 (1994)]
  [hep-th/9407087].

  \bibitem{Seiberg:1994aj} 
  N.~Seiberg and E.~Witten,
  Nucl.\ Phys.\ B {\bf 431}, 484 (1994)
  [hep-th/9408099].

  \bibitem{Shifman:2013ewa} 
  M.~Shifman and A.~Yung,
  arXiv:1304.0822 [hep-th];
  Phys.\ Rev.\ D {\bf 83}, 105021 (2011)
  [arXiv:1103.3471 [hep-th]];
  Phys.\ Rev.\ D {\bf 86}, 065003 (2012)
  [arXiv:1204.4164 [hep-th]];
  Phys.\ Rev.\ D {\bf 86}, 025001 (2012)
  [arXiv:1204.4165 [hep-th]];
  arXiv:1303.1449 [hep-th].
  
  \bibitem{HT1}
  A.~Hanany and D.~Tong,
  JHEP {\bf 0307}, 037 (2003)
  [hep-th/0306150].

  \bibitem{ABEKY}
  R.~Auzzi, S.~Bolognesi, J.~Evslin, K.~Konishi and A.~Yung,
  Nucl.\ Phys.\ B {\bf 673}, 187 (2003)
  [hep-th/0307287].

  \bibitem{SYmon}
  M.~Shifman and A.~Yung,
  Phys.\ Rev.\ D {\bf 70}, 045004 (2004)
  [hep-th/0403149].

  \bibitem{HT2}
  A.~Hanany and D.~Tong,
  JHEP {\bf 0404}, 066 (2004)
  [hep-th/0403158].

  \bibitem{Dorey:1998yh} 
  N.~Dorey,
  JHEP {\bf 9811}, 005 (1998)
  [hep-th/9806056].

  \bibitem{Shifman:2006bs} 
  M.~Shifman, A.~Vainshtein and R.~Zwicky,
  J.\ Phys.\ A {\bf 39}, 13005 (2006)
  [hep-th/0602004].
 
  \bibitem{Shifman:2010kr} 
  M.~Shifman and A.~Yung,
  Phys.\ Rev.\ D {\bf 82}, 066006 (2010)
  [arXiv:1005.5264 [hep-th]].

  \bibitem{HaHo}
  A.~Hanany and K.~Hori,
  Nucl.\ Phys.\  B {\bf 513}, 119 (1998)
  [arXiv:hep-th/9707192].
  
  \bibitem{SY1}
  M.~Shifman and A.~Yung,
  Phys.\ Rev.\  D {\bf 77}, 125016 (2008)
  [arXiv:0803.0158 [hep-th]].

  \bibitem{BSY3}
  P.~A.~Bolokhov, M.~Shifman and A.~Yung,
  Phys.\ Rev.\  D {\bf 81}, 065025 (2010)
  [arXiv:0907.2715 [hep-th]].
  
   \bibitem{Edalati}
  M.~Edalati and D.~Tong,
  JHEP {\bf 0705}, 005 (2007)
  [arXiv:hep-th/0703045].
  
  \bibitem{HSZ}
  A.~Hanany, M.~J.~Strassler and A.~Zaffaroni,
  Nucl.\ Phys.\ B {\bf 513}, 87 (1998)
  [hep-th/9707244].

  \bibitem{VY}
  A.~I.~Vainshtein and A.~Yung,
  Nucl.\ Phys.\ B {\bf 614}, 3 (2001)
  [hep-th/0012250].

  
  \bibitem{AdDVecSal}
  A.~D'Adda, A.~C.~Davis, P.~DiVeccia and P.~Salamonson,
  Nucl.\ Phys.\ {\bf B222} 45 (1983).
   
  \bibitem{ChVa}
  S.~Cecotti and C. Vafa,
  Comm. \ Math. \ Phys. \ {\bf 158} 569 (1993).

  \bibitem{W93}
  E.~Witten,
  Nucl.\ Phys.\ B {\bf 403}, 159 (1993)
  [hep-th/9301042].

  
\end{thebibliography}
\end{document}